\documentclass[showpacs,preprint,showkeys,superscriptaddress,amssymb,
               nofootinbib,aps]{revtex4}
\usepackage{array,multirow}
\usepackage{mdwlist}
\usepackage{amsmath}
\usepackage{amsthm}

\usepackage{amssymb}
\usepackage{bm}
\usepackage{blindtext}
\usepackage{enumitem}
\usepackage{graphicx}
\def\ni{\noindent}
\def\be{\begin{equation}}
\def\ee{\end{equation}}

\begin{document}
\title{\Large Dust content solutions for the \\ Alcubierre warp drive
       spacetime}
\author{Osvaldo L. Santos-Pereira}\email{osvald23@gmail.com}
\affiliation{Instituto de F\' isica, Universidade Federal do Rio de
            Janeiro--UFRJ, 21941-972, Rio de Janeiro, RJ, Brazil}
\author{Everton M.\ C.\ Abreu}\email{evertonabreu@ufrrj.br}
\affiliation{Departamento de F\'{i}sica, Universidade Federal Rural do
	Rio de Janeiro--UFRRJ, 23890-971, Serop\'edica, RJ,  Brazil}
\affiliation{Departamento de F\'{i}sica, Universidade Federal de Juiz de
             Fora--UFJF, 36036-330, Juiz de Fora, MG, Brazil}
\affiliation{Programa de P\'os-Gradua\c{c}\~ao Interdisciplinar em
	F\'{\i}sica Aplicada, Instituto de F\'{\i}sica, Universidade
	Federal do Rio de Janeiro-UFRJ, 21941-972, Rio de Janeiro, RJ,
        Brazil}
\author{Marcelo B. Ribeiro}\email{mbr@if.ufrj.br}
\affiliation{Instituto de F\' isica, Universidade Federal do Rio de
            Janeiro--UFRJ, 21941-972, Rio de Janeiro, RJ, Brazil}
\affiliation{Programa de P\'os-Gradua\c{c}\~ao Interdisciplinar em
	F\'{\i}sica Aplicada, Instituto de F\'{\i}sica, Universidade
	Federal do Rio de Janeiro-UFRJ, 21941-972, Rio de Janeiro, RJ,
        Brazil}
\affiliation{Observat\'orio do Valongo, Universidade Federal do Rio
	de Janeiro--UFRJ, 20080-090, Rio de Janeiro, RJ, Brazil}
\date{\today}
\begin{abstract}
\noindent The Alcubierre metric is a spacetime geometry where a
massive particle inside a spacetime distortion, called warp bubble,
is able to travel at velocities arbitrarily higher than the velocity
of light, a feature known as the warp drive. This is a consequence of
general relativity, which allows global superluminal velocities but
restricts local speeds to subluminal ones as required by special
relativity. In this work we solved the Einstein equations for the
Alcubierre warp drive spacetime geometry considering the dust matter
distribution as source, since the Alcubierre metric was not
originally advanced as a solution of the Einstein equations, but as
a spacetime geometry proposed without a source gravity field. We
found out that all Einstein equations solutions of this geometry
containing pressureless dust lead to vacuum solutions. We also
concluded that these solutions connect the Alcubierre metric to
the Burgers equation, which describes shock waves moving through an
inviscid fluid. Our results also indicated that these shock waves
behave as plane waves. 
\end{abstract}

\maketitle

\section{Introduction}

In general relativity it is possible for particles, in a global sense,
to travel with superluminal velocities whereas the light speed
limit is respected inside a local light cone. The Alcubierre warp
drive metric \cite{Alcubierre1994} satisfies this requirement by
basically producing a spacetime distortion, called \textit{warp bubble},
such that a particle would travel inside this bubble contracting the
spacetime in front of it and expanding the spacetime behind it. In
such a geometrical arrangement, the particle travels globally with
superluminal velocity whereas the warp bubble guarantees that locally
the particle's speed remains subluminal. In its original formulation
it was advanced that this warp metric would imply the violation of
energy conditions, as well as reportedly requiring great amounts of
negative energy density.

Following Alcubierre's original work, several efforts were made to
understand the main caveats of the warp drive metric. Ford and Roman
\cite{FordRoman1996} advanced some quantum inequalities and concluded
that large amounts of negative energy would be required to transport
particles with small masses across small distances. Hence, these
authors concluded that prohibitive huge amounts of negative energy
density would be required to create a warp bubble. Using these quantum
inequalities, Pfenning and Ford \cite{Pfenning1997} calculated the
limits necessary for the bubble parameters and energy values necessary
for the viability of the warp drive, concluding then that the energy
required for a warp bubble is ten orders of magnitude greater than the
total mass of the entire visible universe, also negative.

Krasnikov \cite{Krasnikov1998} discussed the possibility of a massive
particle making a round trip between two points in space faster than
a photon, by arguing that this is not possible when reasonable
assumptions for globally hyperbolic spacetimes are made. He discussed
in details some specific spacetime topologies, assuming that, for some
of them, they need tachyons for superluminal travel to occur. He also
conjectured the need for a possible preparation of a specific spacetime
with some devices along the travel path that would be set up previously
to operate when they were needed for the superluminal travel be possible
without tachyons. Such spacetime was named as \textit{Krasnikov tube}
by Ref.\ \cite{EveretRoman1997}. 

Everett and Roman \cite{EveretRoman1997} generalized the metric proposed
by Krasnikov by hypothesizing a tube along the path of the particle
connecting Earth to a distant star. Inside the tube the spacetime is flat,
but the lightcones are opened out in such a way that they allow the
superluminal travel in one direction. One of the problems mentioned in
Ref.\ \cite{EveretRoman1997} is that even though the Krasnikov tube does
not involve closed timelike curves, it is possible to construct a two
way non-overlapping tube system such that it would work as a time
machine. They also demonstrated that the Krasnikov tube needs great
amounts of negative energy density to function. These authors also used
the generalized Krasnikov tube metric to calculate an
\textit{energy-momentum tensor} (EMT) which would be positive in some
specific regions. Further discussions of the metric proposed by Everett
and Roman \cite{EveretRoman1997} were made by Lobo and Crawford
\cite{Lobo2003,Lobo2002}, who discussed in detail the metric and EMT
derived from it, as well as if it is possible to exist superluminal
travel without the weak energy condition violation. The quantum
inequalities, brought from quantum field theory in Ref.\
\cite{EveretRoman1997}, were also discussed.

Further studies on this subject were made by the following authors.
van de Broeck \cite{Broeck1999} showed how a minor modification of the
Alcubierre geometry can reduce the total energy required for the warp
bubble to distort spacetime. He then presented a modification of the
original warp drive metric where the total negative mass would be of
the order of a few solar masses. Natario \cite{Natario2002} argued
that both the expansion and contraction of space for the Alcubierre
warp drive is a matter of choice, and proposed a new version of the
warp drive theory with zero expansion, a choice of spherical
coordinates, and to use the $x$ axis as the polar axis. Lobo and
Visser \cite{LoboVisser2004} argued that for the Alcubierre warp drive
and its version proposed by Natario \cite{Natario2002}, the center of
the bubble must be massless. They introduced a linearized theory for
both approaches and found that even for low velocities the negative
energy stored in the warp fields must be just a significant fraction
of the particle's mass at the center of the warp bubble. White
\cite{White2003,White2011} described how a warp field interferometer
could be implemented at the Advanced Propulsion Physics Laboratory
with the help of the original Alcubierre's ideas \cite{Alcubierre1994}. 

In this paper we investigate some of these issues. Since the
Alcubierre metric was not originally advanced as a solution of the
Einstein equations, but as an ad hoc proposal aimed at allowing
superluminal global speeds for particles, our aim here is to 
investigate if the dust energy-momentum tensor, the simplest source
matter distribution for the Einstein equations, is able to create a
superluminal warp field. We discuss in detail the dust matter
distribution together with the Alcubierre warp drive metric.
For this matter source the solutions of the Einstein equations require
a zero matter density, \textit{i.e.}, vacuum. Nevertheless, the
resulting vacuum solutions connect the warp drive metric to the Burgers
equation and inviscid fluid with shockwaves, in fact plane waves in
the vacuum.

The plan of the paper is as follows. Sec.\ 2 briefly reviews the
Alcubierre warp drive theory, and in Sec.\ 3  the non-zero components
of the Einstein tensor for the Alcubierre warp drive are written and
the energy conditions are discussed. Sec.\ 4 presents the
Einstein equations written in terms of the dust EMT, analyzes the
expressions obtained for the warp drive metric and presents the
results. Sec.\ 5 depicts our conclusions and final remarks. Appendix
I contains a brief description of the Burgers equation.


\section{Warp Drive Geometry}
\renewcommand{\theequation}{2.\arabic{equation}}
\setcounter{equation}{0}

This section reviews the main aspects of the Alcubierre warp
drive spacetime. The geometrical details of its shape function, which
designs the form of the bubble, are presented, as well as its energy
conditions.

\subsection{The Alcubierre warp drive spacetime}

\begin{figure}[ht]
\includegraphics[scale=0.65]{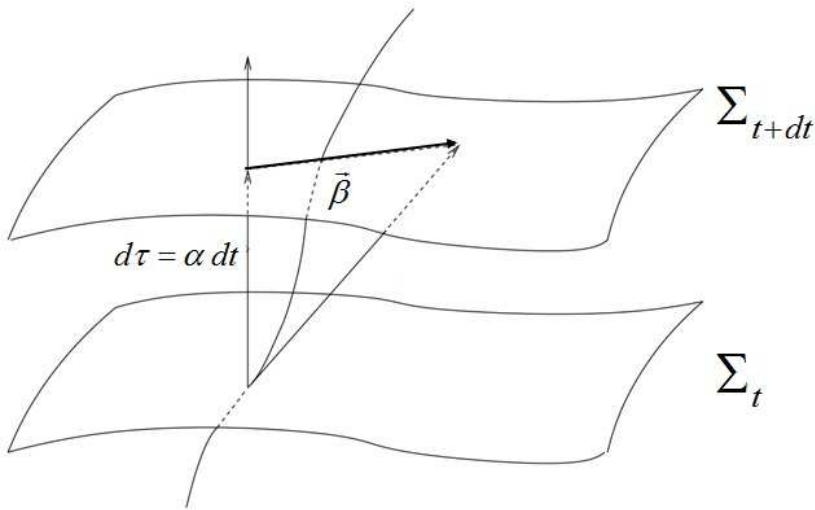}
\caption{Two adjacent spacelike hypersurfaces showing the definitions of
the lapse function $\alpha$ and the shift vector $\beta^i$. It is clear
that the shift vector $\vec{\beta}$ represents how the spacelike
coordinates change from one hypersurface to another as the proper time
elapses. For more details about this 3+1 formalism the reader is refereed
to Ref.\ \cite[Chaps.\ 1-2]{Alcubierre2012} for a clear and concise
explanation.}
\label{adm_1}
\end{figure}

The warp drive geometry \cite{Alcubierre1994} is basically a
spacetime based propulsion system that, in theory, allows a
mass particle to travel with apparent velocities greater than the
light speed by means of a local spacetime distortion that embeds the
particle. The general metric for the warp drive 3+1 formalism
\cite{Alcubierre2012} is given by,
\begin{align}
\nonumber {ds}^2 &= - d\tau^2 = g_{\mu \nu} dx^\mu dx^\nu, \\ 
&= - \left(\alpha^2 -\beta_i\beta^{i}\right) \, dt^2 
+ 2 \beta_i \, dx^i \, dt + \gamma_{ij} \, dx^i \, dx^j,
\label{metric1}
\end{align}
where 
$d\tau$ is the lapse of proper time, $\alpha$ is the lapse function,
$\beta^i$ is the spacelike shift vector and $\gamma_{ij}$ is the
spatial metric for the hypersurfaces.\footnote{From now on Greek indices
will range from 0 to 3, whereas Latin ones indicate the spacelike
hypersurfaces and will range from 1 to 3.} The lapse function $\alpha$
and the shift vector $\beta^i$ are functions to be determined, whereas
$\gamma_{ij}$ is a positive-definite metric on each of the spacelike
hypersurfaces, for all values of time, a feature that makes the
spacetime globally hyperbolic. The lapse of proper time $d\tau$ between
two adjacent hypersurfaces, measured by those observers moving along
the normal direction to the hypersurfaces, also known as Eulerian
observers, is described by the following expression,
\begin{equation}
d\tau = \alpha(t,x^i) dt.
\end{equation}
Fig.\ \ref{adm_1} illustrates the spacetime foliation with two spacelike
hypersurfaces $\Sigma_{t}$ and $\Sigma_{t+dt}$ separated by a timelike
distance $\alpha \, dt$. 

Now, considering the metric (\ref{metric1}), Alcubierre
\cite{Alcubierre1994} assumed the following \textbf{ad hoc} particular
choices for the parameters,
\begin{align}
\alpha &= 1, 
\\
\beta^1& = - v_s(t)f\big[r_s(t)\big], \label{betax}
\\
\beta^2 &= \beta^3 = 0,
\\
\gamma_{ij} &= \delta_{ij}.
\end{align}
Hence, Eq.\ \eqref{metric1} becomes what may be called as the
\textit{Alcubierre warp drive metric}. It may be written as below,
\be
ds^2 = - \left[1 - v_s(t)^2 f(r_s)^2\right]dt^2 - 2 v_s(t) f(r_s)\,dx\,dt 
+ dx^2 + dy^2 + dz^2,
\label{alcmetric1}
\ee
where $v_s(t)$ is the velocity of the center of the bubble moving
along the curve $x_s(t)$. This is given by the following expression,
\begin{equation}
v_s(t) = \frac{dx_s (t)}{dt}.
\end{equation} 

The function $f(r_s)$, named by Alcubierre \cite{Alcubierre1994}
as the warp metric \textit{regulating function}, describes the shape
of the warp bubble. The interior of the bubble is an inertial
reference frame and the observers within it suffer no proper
acceleration. A photon within the warp bubble would always move
faster than a mass particle, as it should according to special
relativity. The regulating function $f(r_s)$ is defined as follows,
\begin{equation}
f(r_s) = \frac{\tanh\left[\sigma(r_s + R)\right] 
- \tanh\left[\sigma(r_s - R)\right]}
{2 \tanh(\sigma R)},
\label{regfunction}
\end{equation}
being, therefore, determined by two arbitrary and positive
parameters, $\sigma$ and $R$. The former is inversely related to
the thickness of the warp bubble, and the latter is proportional to
the bubble's radius. The variable $r_s(t)$ is defined as the distance
from the center of the bubble $[x_s(t),0,0]$ to a generic point
$(x,y,z)$ on the surface of the bubble, as the following expression
shows,
\begin{equation}
r_s(t) = \sqrt{\left[x - x_s(t)\right]^2 + y^2 + z^2}.
\end{equation} 

If one substitutes $x = x_s(t)$ in the warp drive metric in Eq.\
\eqref{alcmetric1} it is straightforward to show that a particle
inside the bubble moves on a geodesic (see Ref.\ \cite{Alcubierre1994})
regardless of the value of $v_s(t)$. Hence, the geodesic $x = x_s(t)$
is physically interpreted as the mass particle trajectory having
no time dilatation, as can be readily seen by the fact that $d\tau
= \alpha dt$, and by the choice of $\alpha = 1$.

The regulating function can be approximated by a step function,
because for distances where $|r_s(t)| < R$ then $f(r_s) = 1$,
whereas for distances where $|r_s(t)| \gg R$ then $f(r_s)
\rightarrow 0$. Hence,
\begin{equation}
    \lim_{\sigma \to \infty} f(r_s) =
    		\begin{cases}
        1, & \text{for} \ \ r_s  \in [-R,R]\\
        0, & \text{otherwise}. 
        \end{cases}
\label{tophat}
\end{equation}
Note that when the regulating function is equal to zero, that is,
outside the warp bubble, according to Eq.\ \eqref{alcmetric1} the
spacetime is flat.

\subsection{Caveats and main points}

Ref.\ \cite{Alcubierre1994} emphasized some fundamental concepts of
special and general relativity theories by arguing that the main
theoretical point behind the warp drive concept is that general
relativity (GR) does not forbid superluminal velocities in a global
sense because spacetime is dynamic. In the 3+1 formalism the
foliation of hyperspace with a time parameter is a perfect tool to
allow a clear interpretation of the results and to prevent the use
of closed causal curves. The warp drive metric relies on the
regulating function $f(r_s)$ to describe the form of the warp bubble.

A simple thought experiment can be advanced in order to demonstrate
that it is possible for particles to make round trips with superluminal
velocity. A particle moving in a local light cone can make a round
trip between points A and B separated by a distance $D$ in a time less
than $2D/c$ measured by an observer that remains always at the place
of departure by just using contraction and expansion of spacetime.
Alcubierre also stated \cite{Alcubierre1994} that the warp drive is
possible without the use of non trivial topologies, such as wormholes.
The particle trajectories are always a timelike curve, regardless of
the parameters used in the Alcubierre metric. The proper time of
distant observers in a flat region is equal to the coordinate time.
This implies that the particle suffers no time dilatation as it moves
on a geodesic.

One of the major issues that seemed to imply the unphysical nature of
superluminal velocities by means of the Alcubierre geometry came from
the initial perception that the warp drive metric violates the three
energy conditions: weak, dominant and strong. By using Eulerian
observers and the warp drive metric the calculations showed that the
energy density becomes everywhere negative \cite{Alcubierre1994}, a
result that implies in the violation of both the weak and dominant
energy conditions. However, this result did not come from the formal
solutions of the Einstein equations, but by finding the Einstein
tensor for the warp drive metric and contracting it to the
4-velocity. By applying the weak energy condition Ref.\ 
\cite{Alcubierre1994} showed that the energy density must be
negative. In other words, similarly to wormholes, superluminal speeds
would require exotic matter. However, such a requirement does not
necessarily eliminate the possibility of using a spacetime distortion
mechanism for achieving a warp drive propulsion system, that is, a
hyper-fast interstellar travel, because Ref.\ \cite{Alcubierre1994}
claimed that although exotic matter may be forbidden classically,
quantum field theory permits the existence of regions with negative 
energy densities, this being the case of the Casimir effect
\cite{DeWitt1979}. 
	
Finally, spacetime topology means that the spacetime described by
the Alcubierre metric is globally hyperbolic. It is then possible
to construct a spacetime that contains closed causal curves using
an idea similar to the one used in Ref.\ \cite{Alcubierre1994}.

\section{The Einstein tensor}

As mentioned above, the Alcubierre metric was not proposed
as a solution of the Einstein field equations, but simply as a
metric whose properties are equivalent to a propulsion system that
drives a mass particle at superluminal global speeds by ``warping''
the spacetime, that is, by generating warp speeds. Hence, the
question that may be posed is what kind of matter or field sources
would be able to produce such spacetime properties. To follow this
analytical path one should couple the Alcubierre metric to the
Einstein field equations in order to try to solve the resulting
equations and draw some conclusions.

\subsection{Einstein tensor components}

We shall start by adopting Alcubierre's original notation of
letting $\beta =-\beta^1=v_s(t) f(r_s)$ in Eq.\ \eqref{betax}, since
the other shift vectors are zero. The components of the Einstein
tensor without a cosmological constant for the warp drive metric
(\ref{alcmetric1}) are given by the expressions below,
\begin{eqnarray} 
G_{00} &=&  - \frac{1}{4} 
(1 + 3\beta^2)
\left[
\left(\frac{\partial \beta}{\partial y} \right)^2 +  
\left(\frac{\partial \beta}{\partial z} \right)^2 
\right] 
- \beta \left(\frac{\partial^2 \beta}{\partial y^2} + 
\frac{\partial^2 \beta}{\partial z^2}\right),
\label{et00}
\\[2pt]
G_{01} &=&  \frac{3}{4} 
\beta \left[
\left(\frac{\partial \beta}{\partial y}\right)^2 
+ \left(\frac{\partial \beta}{\partial z}\right)^2 
\right] 
+ \frac{1}{2}\left(
\frac{\partial^2 \beta}{\partial y^2} 
+ \frac{\partial^2 \beta}{\partial z^2}
\right),
\label{et01}
\\[2pt]
G_{02} &=& - \frac{1}{2}
\frac{\partial^2 \beta}{\partial x \partial y} 
- \frac{\beta}{2} 
\left(2\frac{\partial \beta}{\partial y}
\, \frac{\partial \beta}{\partial x} +
\beta \frac{\partial^2 \beta}{\partial x \partial y} +
\frac{\partial^2 \beta}{\partial t \partial y}\right),
\label{et02}
\\[2pt]
G_{03} &=& - \frac{1}{2}
\frac{\partial^2 \beta}{\partial x \partial z} 
- \frac{\beta}{2} 
\left(2\frac{\partial \beta}{\partial z}
\, \frac{\partial \beta}{\partial x} +
\beta \frac{\partial^2 \beta}{\partial x \partial z} +
\frac{\partial^2 \beta}{\partial t \partial z}\right),
\label{et03}
\\[2pt] 
G_{11} &=& - \frac{3}{4} \left[
\left(\frac{\partial \beta}{\partial y}\right)^2 
+ \left(\frac{\partial \beta}{\partial z}\right)^2
\right], \label{et11}
\\[2pt] 
G_{12} &=& \frac{1}{2}\left(
2 \frac{\partial \beta}{\partial y} \, 
\frac{\partial \beta}{\partial x} 
+ \beta \frac{\partial^2 \beta}{\partial x \partial y} 
+ \frac{\partial^2 \beta}{\partial t \partial y}\right),
\label{et12}
\\[2pt] 
G_{13} &=& \frac{1}{2}\left(
2 \frac{\partial \beta}{\partial z} \, 
\frac{\partial \beta}{\partial x} 
+ \beta \frac{\partial^2 \beta}{\partial x \partial z} 
+ \frac{\partial^2 \beta}{\partial t \partial z}\right),
\label{et13}
\\[2pt] 
G_{23} &=& \frac{1}{2} \frac{\partial \beta}{\partial z} 
\, \frac{\partial \beta}{\partial y},
\label{et23}
\\[2pt] 
G_{22} &=& - \left[
\frac{\partial^2 \beta}{\partial t \partial x}
+ \beta \frac{\partial^2 \beta}{\partial x^2}
+ \left(\frac{\partial \beta}{\partial x}\right)^2
\right]
- \frac{1}{4}\left[
\left(\frac{\partial \beta}{\partial y}\right)^2
- \left(\frac{\partial \beta}{\partial z}\right)^2
\right],
\label{et22}
\\[2pt] 
G_{33} &=& - \left[
\frac{\partial^2 \beta}{\partial t \partial x}
+ \beta \frac{\partial^2 \beta}{\partial x^2}
+ \left(\frac{\partial \beta}{\partial x}\right)^2
\right]
+ \frac{1}{4}\left[
\left(\frac{\partial \beta}{\partial y}\right)^2
- \left(\frac{\partial \beta}{\partial z}\right)^2
\right]\,\,.
\label{et33}
\end{eqnarray}

\subsection{Energy conditions}

The components for the Eulerian (normal) observers' 4-velocities are
given by,
\begin{equation}
u^\alpha = \left[1, - v_s(t) f(r_s), 0, 0\right], \ \ 
u_\alpha = (- 1,0,0,0).
\label{eulerianvel}
\end{equation}
Using, as below, these results in the Einstein equations,
\begin{equation}
T_{\alpha \beta} u^\alpha u^\beta = \frac{1}{8\pi}G_{\alpha \beta} 
u^\alpha u^\beta\,\,,
\label{eq22}
\end{equation} 
allows us to obtain an expression concerning the energy conditions.
Considering Eqs.\,\eqref{eulerianvel} and that the only non-zero terms
of Eq.\ (\ref{eq22}) are $G_{00}$, $G_{01}$ and $G_{11}$, we obtain
the following expression,
\begin{equation}
T_{\alpha \beta} \, u^\alpha u^\beta 
= \frac{1}{8\pi}\left(G_{00} - 2 v_s f G_{01} 
+ v_s^2 f^2 G_{11}\right).
\label{endenalc}
\end{equation}
Substituting Eqs.\,\eqref{et00}, \,\eqref{et01} and \eqref{et11} into
Eq.\ \eqref{endenalc} the result may be written as below,
\begin{equation}
T_{\alpha \beta} \, u^\alpha u^\beta 
= - \frac{v_s^2}{32 \pi}
\left[
\left(\frac{\partial f}{\partial y} 
\right)^2 +  \left(\frac{\partial f}{\partial z} 
\right)^2
\right].
\label{edwd1}
\end{equation}
This expression can be physically interpreted as being a matter-energy
density as observed in the frame of Eulerian observers. Besides, since
the bubble radius is given by, 
\begin{equation}
r_s = \sqrt{(x - x_s)^2 + y^2 + z^2},    
\label{eq23}
\end{equation}
applying implicit partial derivative rules, Eq.\ (\ref{edwd1})
takes then the form below,
\begin{equation}
T_{\alpha \beta} u^\alpha u^\beta  
= - \frac{v_s^2}{16 \pi}\frac{y^2 + z^2}{r_s^2}
\left(\frac{\partial f}{\partial r_s}\right)^2 \leq 0.
\label{edwd2}
\end{equation}

The expression above shows that the energy density can only
vanish or assume negative values, a result that Ref.\
\cite{Alcubierre1994} assumed as being a necessary condition
for exotic matter and faster than light travel. In addition,
Ref.\ \cite{Alcubierre1994} also emphasized that this result
violates the energy conditions because the energy density
becomes everywhere negative. There are four energy conditions
in general relativity \cite{HawkingEllis1973} and, according
to Alcubierre, the warp drive metric violates both the weak
and dominant ones. 

Nevertheless, it must be noted that although Alcubierre stated
\cite[Eq.\ 19]{Alcubierre1994} that the relation $T_{\alpha
\beta} u^\alpha u^\beta$ must be everywhere negative, the weak,
strong, null, and dominant energy conditions can still be satisfied
if this contraction is equal to zero \cite{HawkingEllis1973}.


\section{Dust content energy-momentum tensors}
\renewcommand{\theequation}{3.\arabic{equation}}
\setcounter{equation}{0}

In this section we shall discuss matter content solutions of the
Einstein's equations considering the dust EMT for the Alcubierre 
metric. This is the simplest possible matter content that
can be studied as possible source for warp speeds. The dust solution
contains only matter-energy density and depends on the 4-velocities
of the observables $T_{\alpha \beta}=\mu u_\alpha u_\beta$. From
now on we shall assume the following form for the Einstein field
equations,
\begin{equation}
G_{\mu\nu} = 8 \pi T_{\mu\nu}.    
\label{eins-eq}
\end{equation}

\subsection{Dust warp metric solutions}

The dust solution is an exact solution of the Einstein equations for
fluids where gravity is produced by the mass density of pressureless
particles. It can be understood as a model for a configuration of dust
particles that move with gravity alone and, hence, there is no other
type of interaction among them. This solution is used to model
gravitational collapse, as well as in cosmology, since galaxies are
considered the basic building blocks of the universe whose main
interaction is due to the general geometrical background. One can
further envisage a possible interest in this solution if one
considers the galactic disks as being modeled by finite rotating
disks of dust.

The stress-energy tensor of a relativistic fluid with no pressure can 
be written in the simple form below,
\begin{equation}
T_{\alpha \beta} = \mu \, u_{\alpha} u_{\beta},
\end{equation}
where $\mu$ is a scalar function that represents the matter density. 
Considering Eqs.\ (\ref{eulerianvel}), the stress-energy tensor yields,
\begin{equation}
T_{\alpha\beta} = 
\begin{pmatrix} 
\mu & 0 & 0 & 0  \\ 
0   & 0 & 0 & 0  \\ 
0   & 0 & 0 & 0  \\ 
0   & 0 & 0 & 0 
\end{pmatrix}.
\label{dustemt}
\end{equation}

In order to solve the Einstein equations one has to use the tensor
components given in Eqs.\ \eqref{et00} to \eqref{et33} with the dust
EMT above. Substituting $G_{11} = 8\pi T_{11}$ and $G_{01} = 8 \pi
T_{01}$ from the Einstein equations into the component $G_{00} = 8
\pi T_{00}$, the resulting expression may be written as follows,
\begin{equation}
T_{00} + 2 \beta T_{01} + \frac{1}{3}(3\beta^2 - 1)T_{11} = 0\,\,.
\label{matterzero}
\end{equation}
Now, considering Eq.\ \eqref{dustemt} in the equation above implies
in a vanishing matter density,
\begin{equation}
\mu = 0.
\end{equation}

This result implies that a warp bubble cannot be created with a dust
matter distribution as source. Nevertheless, the other components of
the Einstein equations lead to some interesting features for the
Alcubierre warp drive metric, as we shall see below.

Since $T_{23} = 0$, the result below follows from equation
$G_{23} = 8 \pi T_{23}$,
\begin{equation}
G_{23} = \frac{1}{2} \frac{\partial \beta}{\partial z}
\, \frac{\partial \beta}{\partial y}  = 0,
\end{equation}
which means that either $\partial \beta/ \partial z$, or $\partial
\beta/\partial y$, or both, vanish. Let us now discuss both cases.

\begin{description}[align=left]
\item[Case 1:\small $\bm{\left[\displaystyle \frac{\partial \beta}
	{\partial z}=0\right]}$]
     This means that the function $\beta $ does not depend on the
     coordinate $z$ and the Einstein tensor components $G_{13}$,
     $G_{03}$ and $G_{23}$ are identically zero. Substituting
     $\partial \beta/\partial z = 0$ into the component $G_{11} = 8
     \pi T_{11}$, and since $T_{11} = 0$, it follows immediately that,
     \begin{equation}
        \frac{\partial \beta}{\partial y}  = 0,
     \end{equation}
     which means that the function $\beta$ does not depend on the
     $y$-coordinate either. From Eq. \eqref{dustemt} it is
     straightforward to verify that the components $G_{22} = 8 \pi
     T_{22}$, and $G_{33} = 8 \pi T_{33}$ are also zero. Therefore,
     the field equations are reduced to,
     \begin{equation}
	\mu = 0,
	\label{mu0}
     \end{equation}
     \begin{equation}
       \frac{\partial^2 \beta}{\partial t \partial x}
       + \beta \frac{\partial^2 \beta}{\partial x^2}
       + \left(\frac{\partial \beta}{\partial x}\right)^2 = 0.
     \label{advec1}
     \end{equation}

\item[Case 2:\small$\bm{\left[\displaystyle \frac{\partial \beta}
	{\partial y}=0\right]}$] 
     This means that the function $\beta$ does not depend on the 
     $y$-coordinate and, consequently, the Einstein tensor components
     $G_{12}$, $G_{23}$ and $G_{02}$ are identically zero. Since
     $G_{11} = 8 \pi T_{11}$, and $T_{11} = 0$, it follows immediately
     that
     \begin{equation}
	     \frac{\partial \beta}{\partial z}  = 0,
     \end{equation}
     which means that the function $\beta$ does not depend on the
     $z$-coordinate either. Hence, the set of field equations are
     also reduced to the expressions (\ref{mu0}) and (\ref{advec1}).
\end{description}
\vspace{-0.6cm} $\square$ \vspace{0.3cm}

Both cases above lead to the same results, so it does not matter
if $\partial \beta/\partial y = 0$ or $\partial \beta/\partial z = 0$.
In addition, they both lead to a vanishing matter density $\mu = 0$
and, consequently, the energy density found in Eqs.\ (\ref{edwd1})
and (\ref{edwd2}) must be zero. This means that the energy conditions
in Eqs.\ (\ref{edwd2}) are immediately, and trivially, satisfied. One
must mention that such trivial result is a consequence of the dust
case EMT in the Alcubierre warp drive metric leading back to vacuum
solution, which might not necessarily happens when one considers
more complex dust or energy content EMTs.

Nevertheless, one is still left with a single partial differential
equation to solve, Eq.$\,$(\ref{advec1}), which can be rewritten as
below, 
\begin{equation}
\frac{\partial}{\partial x}\left[
\frac{\partial \beta}{\partial t}
+ \frac{1}{2} \frac{\partial}{\partial x} (\beta^2)
\right]
= 0\,\,,  
\label{advec3}
\end{equation}
whose integration is straightforward if we remember that
$\beta=\beta(t,x)$, yielding,
\begin{equation}
\frac{\partial \beta}{\partial t}
+ \frac{1}{2}\frac{\partial}{\partial x} (\beta^2)
= h(t),   
\label{advec4}
\end{equation}
where $h = h(t)$ is an arbitrary function of $t$ to be
determined by boundary conditions.

In its homogeneous form, that is, for $h(t) = 0$, Eq.\,(\ref{advec4})
becomes the conservative form of the \textit{inviscid Burgers
equation} (see Appendix I for details), a well known equation
appearing in fluid models, such as gas dynamics and traffic flows, as
well as in hyperbolic equations and conservation laws. It is a
quasilinear hyperbolic equation and its current density is the kinetic
energy density. If one defines the flow density as being given by $J_f
= J_f(\beta)$, which can be a general function of $\beta$, and let it
be given by the following expression,
\begin{equation}
J_f(\beta) = \beta^2,    
\end{equation}
Eq.\ \eqref{advec4} may be rewritten as below, 
\begin{equation}
\frac{\partial \beta}{\partial t}
+ \frac{1}{2} \frac{\partial}{\partial x} J_f
= h(t).  
\label{eq511}
\end{equation}
The phenomena arising from the Burgers equation are conservation laws
and the formation of shock waves, that is, discontinuities that appear
after a finite time and then propagate in a regular manner. The one
dimension conservation law implicit in Eq.\ \eqref{eq511} can be seen
if we note that equations of the following form,
\begin{equation}
\frac{\partial u}{\partial t} + 
\frac{\partial}{\partial x}F(u) = 0,
\label{bur1}
\end{equation}
can be interpreted as a conservation law \cite{Evans2010}, where the
function $u = u(t,x)$ is to be determined with the initial condition
below,
\begin{equation}
u(t=0,x) = u_0(x).
\end{equation} 

The Burgers equation can describe rarefaction and expansion waves.
Hence, in the present context Eq.\ (\ref{eq511}) can depict a
spacetime shock wave, in this case as a plane wave. It is worth
noticing that Alcubierre built a warp drive bubble in the vacuum,
so when we impose the dust solution to the EMT and solve the
Einstein equations for the warp drive metric the matter density
vanishes, recovering the vacuum, but also showing that the warp
bubble regulating function may obey the inviscid Burgers equation in
the particular case when the function $h(t)$ vanishes.

In the warp drive scenario, $\beta = v_s(t) f(r_s)$ would
be interpreted as a boost in the $x$-direction, which means that the
warp bubble obeying the more general Burgers equation (\ref{advec4})
the warp drive may then be understood as conservation of linear
momentum in the $x$-direction when $h(t)= \text{constant}$. Then
$\partial \beta/ \partial t$ may be interpreted as a force per unit
mass, \textit{i.e.}, the time derivative of momentum, and
$(1/2)\partial (\beta^2)/\partial x$ would be a potential,
\textit{i.e.}, the divergence of the total energy that is entirely
kinetic. This would seem reasonable, since the EMT for the dust
solution implies no interaction among particles, so the self
gravitating potential is neglected. Note that both cases 1 and 2
above lead to the same results, being then a consequence of the
symmetric properties of both the Einstein equations and EMTs. Table
\ref{tab1} summarizes the results obtained above.
\begin{table}[h!]
\begin{tabular}{| m{3cm} | m{3cm} | m{5cm} |}
\hline 
Case & Consequence & Results \\ 
\hline 
$1) \ \displaystyle{\frac{\partial \beta}{\partial z} = 0} $ 
& 
$\displaystyle{\frac{\partial \beta}{\partial y} = 0}$
&
$\begin{array} {ll} 
\mu = 0 \\ [6pt]
\beta = \beta(t,x) \\ [6pt]
\displaystyle{\frac{\partial \beta}{\partial t} 
+ \frac{1}{2} \frac{\partial}{\partial x}(\beta^2)
= h(t)}
\\[10pt]
\end{array}$ \\ 
\hline 
$2) \ \displaystyle{\frac{\partial \beta}{\partial y} = 0}$
&
$\displaystyle{\frac{\partial \beta}{\partial z} = 0}$ 
& 
$\begin{array} {ll} 
\mu = 0 \\ [6pt]
\beta = \beta(t,x)\\ [6pt]
\displaystyle{\frac{\partial \beta}{\partial t} 
+ \frac{1}{2} \frac{\partial}{\partial x}(\beta^2)
= h(t)}
\\[10pt]
\end{array}$ \\
\hline 
\end{tabular}
\caption{Summary of results for the warp drive spacetime having dust
matter content.}
\label{tab1}
\end{table}

\subsection{The warp metric and shock waves} \label{shockwave}

We have seen that the Burgers equation appeared here as a vacuum
solution of the warp drive metric when one considers this geometry
in the Einstein equations with dust matter source. In recap, the
current density was given by $\beta = v_s(t) f(r_s)$, where $v_s$
is the bubble velocity, $f(r_s)$ is the regulating function of the
bubble shape and the inviscid Burgers equation \eqref{bur1}
represents a conservation law for this current density. This result
can be physically understood as a conservation law. Analyzing each
term of the Burgers equation as it was seen in Eq.\eqref{advec4},
the first term in the left hand side, $\partial \beta/\partial t$,
can be interpreted as a type of force per unit mass, \textit{i.e.},
the time derivative of momentum, since for the warp drive $\beta=
v_s(t) f(r_s)$ and this function contains the bubble velocity and
shape in the $x$-direction. The second term on the left hand side,
$\frac{1}{2}\,\partial (\beta^2)/\partial x$, can be understood as
the divergence of the total energy, which is entirely kinetic. The
right hand side is a function $h(t)$ of the time coordinate only,
which can be determined by boundary conditions. When $h(t) = 0$ 
the inviscid Burgers equation is recovered, which is a kind of
conservation equation. Physically it can be understood considering
the warp metric as a conservation of both the energy and momentum
in the direction of the wave propagation. 

The dust solutions lead to zero matter density, and the Burgers equation
arose in this context as a regulating function parameter. This result
suggests that the necessary energy to create the associate shock wave
is purely geometrical. The fact that the Burgers equation appeared as
part of the solution of the Einstein equations relative to the vacuum
solution is a very interesting evidence that the warp drive metric can
be understood as spacetime motion equivalent to a shock wave moving in
a fluid.

As an analogy, shock waves are produced when, for example, an aircraft
traveling at high subsonic velocity produces sound waves that piled up
due to the air surrounding the aircraft traveling at local speed of
sound, causing a kind of explosion. This result may be considered
intuitive for the perfect fluid, but it is not clear if this can be
the case for dust because, as seen above, in this case it happens in
the vacuum.

\subsection{Divergence for the Dust EMT} \label{divemts}

Calculating the divergence for the dust EMT, and demanding that it 
should be null, one arrives at the following condition,
\begin{equation}
\mu \frac{\partial \beta}{\partial x} = 0,
\label{nulldiv}
\end{equation}
which is immediately satisfied since the matter density vanishes 
for the dust EMT.


\section{Conclusions}
\renewcommand{\theequation}{5.\arabic{equation}}
\setcounter{equation}{0}

In this work we have analyzed the solutions of the Einstein equations
for the Alcubierre warp drive spacetime with the choice of dust
energy-momentum tensor (EMT) as possible source of global
superluminal particle velocities, that is, warp speeds. The Einstein
equations are reduced to two solutions that lead to the same results.
The matter density becomes equal to zero, $\beta(t,x)$ becomes a
function of the time and $x$ coordinates only and the Burgers equation
appears as a special case of the vacuum warp drive spacetime.
The divergence for the dust EMT is zero. In addition, all the
energy conditions are trivially satisfied.

Summing up, we showed that if one starts with dust only EMT 
the Alcubierre type warp drive is not possible, since the Einstein
equations lead the solutions back to vacuum. A more complex matter
distribution source for the EMT than simple dust is possibly required
for a warp drive bubble including, perhaps, some eletromagnetic
components in the EMT or even including the cosmological constant in
the Einstein equations. These issues are the subject of ongoing
research.

Regarding Eq.\,\eqref{advec4}, two possibilities are worth considering.
If $h(t)=0$ the inviscid Burgers equation is recovered, which is a type
of a conservation equation with discontinuities, that is, shock waves,
in the case studied here as plane waves. If $h(t)$ were to be of the
form
\begin{equation}
h(t) = \nu \frac{\partial^2 \beta}{\partial x^2},
\label{advecburger}
\end{equation}
then Eq.\ \eqref{advec4} becomes the viscous Burgers equation, valid
for a dissipative system, where $\nu$ is a diffusion coefficient. 
It must also be mentioned that the Burgers equation is found in other
relativistic solutions as well, such as FLRW cosmology and
Schwarzschild background \cite{turcos}. Based on Eq.\
\eqref{advecburger}, one might speculate that $h(t)$ \textbf{could,
perhaps,} acts as a source term that originates the shock waves.


\section*{Acknowledgments}

\ni We are grateful to an anonymous referee for important insights,
suggestions and recommendations that improved the paper. E.M.C.A.\
thanks CNPq (Conselho Nacional de Desenvolvimento Cient\'ifico e
Tecnol\'ogico), Brazil's government scientific supporting agency, for
partial financial support, grants numbers 406894/2018-3 and 302155/2015-5. 


\section*{Appendix I: The Burgers equation}
\renewcommand{\theequation}{I.\arabic{equation}}
\setcounter{equation}{0}

The Burgers equation is a very famous nonlinear partial differential
equation due to its application in areas such as fluid and gas dynamics,
traffic flow, acoustics, shock waves, and so forth. The equation was
first discovered by Forsyth \cite{Forsyth1906} and later by Bateman
\cite{Bateman1915}. However, the equation was named after Burgers
\cite{Burgers1948} because of his extensive work upon the issue. The
general form of the viscous Burgers equation in one space dimension is  
\begin{equation}
\frac{\partial u}{\partial t} + u\frac{\partial u}
{\partial x} = \nu \frac{\partial^2 u}{\partial x^2}\,\,,
\end{equation}
where $u(x,t)$ is a field dependent of both time $t$ and the
$x$-coordinate, and $\nu = \nu(u,x,t)$ is the diffusion coefficient. 
In this form it is known from its use in modeling dissipative systems. 
When the diffusion term is zero, $\nu = 0$, the Burgers equation 
assumes its inviscid form, which means that the equation is free of 
the viscosity term, 
\begin{equation}
\frac{\partial u}{\partial t} + 
u\frac{\partial u}{\partial x} = 0\,\,.
\label{burginvisc1}
\end{equation}
This can also be rewritten in its conservative form below
follows, 
\begin{equation}
\frac{\partial u}{\partial t} + 
\frac{1}{2}\frac{\partial}{\partial x}(u^2) = 0\,\,,
\label{burginvisc2}
\end{equation}

The inviscid form of the Burgers equation can be classified as 
a quasilinear hyperbolic equation, used as a model for a 
conservation equation. The solution of the inviscid Burgers 
equation can be constructed by using the characteristics method,   
namely, for an initial condition $u(x,0) = g(x)$, the solution 
is given by a wave solution of the form $u(x,t) = g(x - ut)$.
The interested reader will find in Ref.\ \cite{Evans2010} a
discussion about the explicit solution of the inviscid Burgers
equation.

The solution $u(x,t) = g(x - ut)$ is written in such a way 
that its characteristics do not intersect each other. Besides, 
when they do intersect, the inviscid Burgers equation leads us 
to a shock waves framework, which is deemed as a propagation 
of a perturbation. See Ref.\ \cite{Cole1951} for
more details on shock wave theory in viscous fluids and how it
relates to turbulence theory.

When one thinks about waves in fluids, the 
shock waves appear when the waves moves faster than the local 
velocity of sound in this very fluid. They can be characterized 
by an abrupt modification of the pressure, temperature and 
density of the medium. Shock waves velocity and energy dissipate 
quickly as a function of the distance. Another interesting 
feature concerning shock waves is that they keep the energy, 
but they increase the entropy of the system \cite{Evans2010}. 
Consequently, the decrease in energy for a shock wave can be 
transformed into work in order to keep energy for the system. 

In the warp drive scenario, when the dust EMT solution was proposed,
the current density of the Burgers equation appeared as ${v_s}^2 f^2$.
Hence, it can be understood as the conservation of the kinetic energy 
field of the warp bubble.


\end{document}